\documentclass[aps,pra,twocolumn,showpacs,amsmath,amssymb,groupaddress,longbibliography]{revtex4-2}
\usepackage[utf8]{inputenc}
\usepackage{graphicx,graphics}
\usepackage[caption=false]{subfig}
\usepackage{hyperref}
\hypersetup{
  colorlinks=true,linkcolor=blue,urlcolor=blue,citecolor=blue}

\newcommand{\br}{\mathrm{\mathbf{r}}}
\newcommand{\bk}{\mathrm{\mathbf{k}}}
\newcommand{\ha}{\hat{a}}
\newcommand{\hb}{\hat{b}}
\newcommand{\hPsi}{\hat{\Psi}}
\newcommand{\sgn}{\mathrm{sgn}}

\begin{document}
\title{Theoretical Exploration of Phase Transitions in a Cavity-BEC
  System with Two Crossed Optical Pumps}

\author{Wei Qin}
\altaffiliation{These authors contributed equally to this work.}
\author{Dong-Chen Zheng}
\altaffiliation{These authors contributed equally to this work.}
\author{Zhao-Di Wu}
\author{Yuan-Hong Chen}
\author{Renyuan Liao}
\email{ryliao@fjnu.edu.cn}
\affiliation{Fujian Provincial Key Laboratory for Quantum Manipulation and New Energy Materials, College of Physics and Energy, Fujian Normal University, Fuzhou 350117, China}
\affiliation{Fujian Provincial Collaborative Innovation Center for Advanced High-Field Superconducting Materials and Engineering, Fuzhou, 350117, China}
\date{\today}
\begin{abstract}
   We consider a Bose-Einstein condensation (BEC) inside an optical cavity and two crossed coherent pump fields. We determine the phase boundary
separating the normal superfluid phase and the superradiance phase, perturbatively. In the regime of negative cavity detuning, we map out the phase diagrams both for  an attractive and a repulsive optical lattice. It turns out that the situation is quite different in two cases. Specifically, in the case of attractive lattice, if a system is in the superradiant phase with one pump laser, adding another pump does not drive the system out of the superradiance phase. While for the repulsive lattice, increasing another pump potential have suppressive effects on the superradiance. We also find that, in the case of attractive lattice, equally increasing two pump lattice potentials can induce a transition from the normal phase to the superradiance phase. In stark contrast, for the repulsive lattice, the system will remain in the normal phase as the  pump depths are tuned within a wide range, independent of the cavity detuning and the decay rate.
\end{abstract}
\maketitle

\section{Introduction}

Loading ultracold quantum gases into single or multiple high-finesse cavities provides a versatile platform to explore many-body phenomena~\cite{ESS13,LEV182,RIT21,MIV21,HEM22,DON22,YEL22,ESS23,COS23,DEN23,BRA23}.The coupling between the degenerate quantum gases and quantized radiation fields in cavities gives rise to controllable long-range interactions, which is dominant in many-body phases~\cite{ESS10,RMP23}. Transversely driving a Bose-Einstein condensation (BEC) inside an optical cavity is such a typical example, and it allows to achieve the Dicke model and a self-organization phase~\cite{DIC54,EPJD08,EPJD082,ESS10,DOM10,ESS11,BAR12,ZWE13,HEM15}. Taking the atoms' spins into account, one single cavity can also
mediate spin-dependent interactions, realize single-mode Dicke spin-models~\cite{BAR17,ESS18,LEV18,BAR18}, and simulate a nondegenerate two-mode Dicke model~\cite{ESS19,ESS21}. With two crossed linear cavities, it is possible to demonstrate the coupling between two order parameters~\cite{SAC11,FRA15,ESS182}, realize different supersolid formation, and even engineer a continuous $U(1)$ symmetry in real space and cavity space~\cite{DON17,ZWE17,ZHA18,LAN21}, so that Higgs and Goldstone modes can be monitored and manipulated~\cite{TIL17}.

Scattering photons into cavities, companied with the self-organization of quantum gases, lies at the heart of these phenomena, which is also
referred as the superradiance. Usually, an attractive potential, rather than a repulsive one, is expected to induce the superradiance.
Since, intuitively, the buildup of any additional repulsive potential seems to cost energy, and hence prohibits self-organization.
Actually, most of experiments implement a red-detuned pump laser to drive atoms, so that an attractive standing wave lattice can be
generated. However, a repulsive pump potential, generated by a blue-detuned optical pump, is also able to produce the superradiance, which has been verified theoretically and experimentally~\cite{SIM10,JIA11,SIM12,PIA17,DON19}.

In practice, such a cavity-BEC system is commonly driven by a single pump laser. It is hard to resist the temptation to ask: what if there are
more than just one pump? And what is the situation in the case of an attractive or a repulsive optical lattice? It is not a trivial
question to answer since additional pumps produce a two-dimensional periodic potential and lower the symmetry of the system, as compared with the case of one pump.
In this work, we address this question by considering a cavity-BEC system in the presence of two crossed optical pumps, and we also take the interference between the two pumps
into account. By mapping out the phase diagrams, we conclude that adding another pump will increase the tendency of transition towards the superradiance in the case of attractive lattice, and suppress the tendency in the case of repulsive lattice. We also find that the repulsive rectangular lattice (equal pump lattice depths) can not induce the superradiance within a wide range of pump depths, which is independent of cavity detuning and decay rate.

\section{Model and Formalism}

\begin{figure}[!htbp]
  \subfloat
  {\centering 
    \label{fig:Apparatus} \label{fig:K_Dist}
    \label{fig:R_Dist_NP} \label{fig:R_Dist_SR}
    \includegraphics[width=0.47\textwidth]{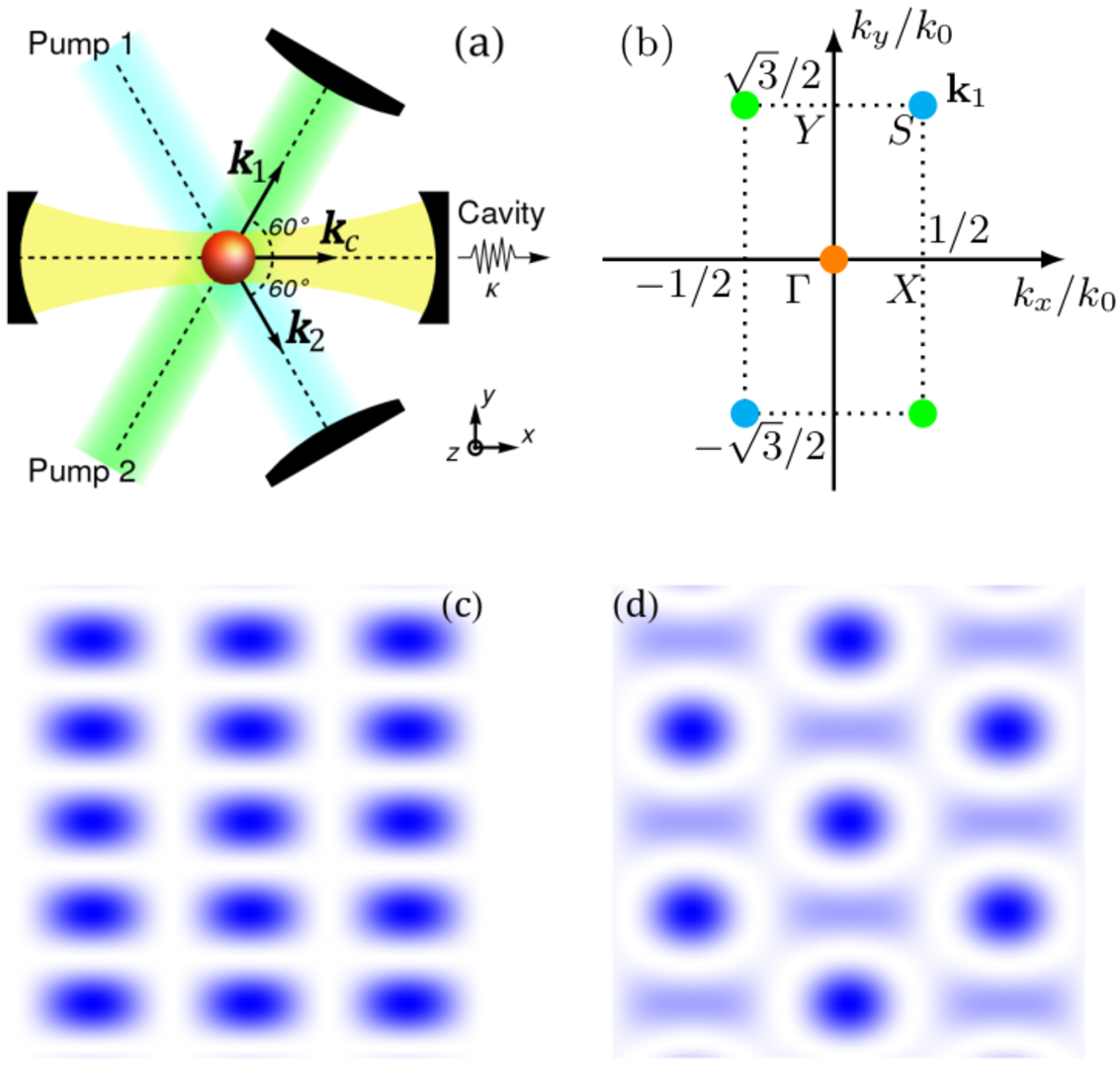}
  }
  \caption{(a) Schematic of the experimental setup. A BEC (orange)
    inside a high-finesse optical resonator is exposed to pump laser
    beam 1 (green) and pump laser beam 2 (blue). Photons scattered by
    the atoms populate the cavity mode (yellow), with coherent field
    amplitude $\alpha$ that can be detected when leaking from the
    cavity. The angles between the cavity beam and per pump beam are
    both $60^{\circ}$.
    (b) Wave vector (momentum) distributions of the BEC. Orange point
    marks the condensed state in absence of photons scattering into
    the cavity, while green (blue) point is steady state of condensation
    scattered by pump 1 (2) in superradiant phase. The dotted line is
    the boundary of the first Brillouin zone for the two dimensional
    rectangular lattice of the combined pump fields.
    (c) and (d) are calculated density modulations for equal pump
    lattice depths without and with the intracavity field,
    respectively, for red atomic detuning $\Delta_{a}<0$.
    In the case of $\Delta_{a}>0$, there is no superradiant phase
    under equal pump depths within red atomic detuning $\Delta_{c}<0$,
    but still has the same density pattern as (c) in absence of
    intracavity field.}
  \label{fig:Schematic_Diagrams}
\end{figure}

We consider a BEC (orange) inside a high-finesse optical Fabry-P$\acute{e}$rot resonator is exposed to a cavity laser beam (yellow) and two pump laser beams (green and blue). The angle between the cavity beam and per pump beam is $60^{\circ}$, as shown in FIG.~$\ref{fig:Schematic_Diagrams}\subref{fig:Apparatus}$. We focus on the two-dimensional case, where the atoms' motion
along $\hat{z}$ can be frozen by tight confinement and bosons can only move in the $xy$ plane~\cite{CHE14}. The decay rate of the cavity field is $\kappa$.
Within dipole and rotating-wave approximation, the effective many-body Hamiltonian in a frame rotating at the pump laser frequency~\cite{ESS13,ESS21} is given by
\begin{subequations}
  \label{eq:Hamiltonian_Origin}
\begin{align}
  \hat{H}
  & = \int d\br \hPsi^{\dag}(\br)
    \left( \frac{\hat{\mathbf{p}}^{2}}{2m}
    + \frac{\hbar}{\Delta_{a}}
    \hat{\mathrm{E}}^{\dag}\hat{\mathrm{E}}
    \right) \hPsi(\br)
    -\hbar\Delta_{c} \ha^{\dag}\ha, \\
  \hat{\mathrm{E}}
  & = \sum_{i=1,2}\Omega_{i} \cos\left(\bk_{i}\cdot\br\right)
        + g\cos\left(\bk_{c}\cdot\br\right) \ha,
\end{align}
\end{subequations}
where $\hPsi^{\dag}$($\hPsi$) is the creation (annihilation) operator for bosonic atoms with mass $m$ and momentum $\hat{\mathbf{p}}=-i\hbar\nabla$.
$\ha^{\dag}$($\ha$) is the photon creation (annihilation) operator for the cavity mode. Here we consider pump beams have the same frequency $\omega_{p}$,
which is far detuned by $\Delta_{a}=\omega_{p}-\omega_{a}$ from the atomic resonance at frequency $\omega_{a}$.
The cavity frequency $\omega_{c}$ is closely detuned by $\Delta_{c}=\omega_{p}-\omega_{c}$ from the pump frequency, and we only stay within $\Delta_{c}<0$ throughout this
work. $\Omega_{j}$ with $j=1,2$ are the Rabi frequencies for pump beams, $g$ is single-photon Rabi frequency of the cavity mode.
The wave vectors of the pump beams and cavity light are given by $\bk_{j}=k_{0}\cos(60^{\circ})\hat{x}+(-1)^{j+1}k_{0}\sin(60^{\circ})\hat{y}$ and $\bk_{c}=k_{0}\hat{x}$ respectively,
as denoted in FIG.~$\ref{fig:Schematic_Diagrams}\subref{fig:Apparatus}$. Here $k_{0}$ is the wave-vector magnitude of the pumping lasers and the cavity mode. We choose recoil energy $E_{R}=\hbar^{2}k_{0}^{2}/2m$ as energy unit, and set $\hbar=1$ for simplicity.

At zero temperature limit, all atoms occupy the lowest-energy
Bloch state $|\Psi^{(1)}(0)\rangle$ in absence of intracavity
photons, where $|\Psi^{(j)}(\bk)\rangle$ denotes the $j$-th band
eigenstates of the single-particle Hamiltonian
\begin{align}
  \hat{\mathcal{H}}_{0}
  &= \frac{\hat{\mathbf{p}}^{2}}{2m}
    + \sgn(\Delta_{a}) \Bigg[
    V_{1}\cos^{2}(\bk_{1}\cdot\br) + V_{2}\cos^{2}(\bk_{2}\cdot\br)
    \nonumber \\
  & \hspace{5em} + 2\sqrt{V_{1}V_{2}}
    \cos(\bk_{1}\cdot\br)\cos(\bk_{2}\cdot\br) \Bigg].
\end{align}
Here we defined pump lattice depths $V_{i}=\Omega_{i}^{2}/|\Delta_{a}|$ with
$i=1,2$, and $\mathrm{sgn}(\Delta_{a})$ gives the sign of atomic
detuning $\Delta_{a}$.
Indeed, $\mathrm{sgn}(\Delta_{a})$ determines whether the effective
pump potential is attractive (negative) or repulsive (positive).

In the presence of intracavity photons, they are scattered by atoms from the transverse pump lasers into the cavity mode and vice versa.
It is sufficient to consider the two-photon scattering processes through the phase transition. Equivalently, starting from the lowest-energy Bloch state
$|\Psi^{(1)}(0)\rangle$ of BEC, only $|\Psi^{(j)}(\bk_{1})\rangle$ can be reached by applying scattering operator
\begin{align}
  \hat{\Theta} = \sum_{i=1,2}
  \sqrt{V_{i}}\cos(\bk_{i}\cdot\br)\cos(\bk_{c}\cdot\br).
\end{align}
Hence the atomic field operator can be expanded as
\begin{align}
  \label{eq:Psi_Expansion}
  \hPsi(\br)
  &\approx \langle\br|\Psi^{(1)}(0)\rangle\hb_{0}
    + \sum_{j} \langle\br|\Psi^{(j)}(\bk_{1})\rangle\hb_{j},
\end{align}
where $\hb_{0}$ and $\hb_{j}$ are bosonic annihilation operators for states $|\Psi^{(1)}(0)\rangle$ and $|\Psi^{(j)}(\bk_{1})\rangle$ respectively.
The sums are over all band index $j=1,2,\dots$, whilst the particle number conservation
$\hb_{0}^{\dag}\hb_{0} + \sum_{j}\hb_{j}^{\dag}\hb_{j}=N$ should be satisfied, with $N$ being the total atom number.
Substituting Eq.~(\ref{eq:Psi_Expansion}) into Eq.~(\ref{eq:Hamiltonian_Origin}) yields the effective Hamiltonian
\begin{align}
  \label{eq:Hamiltonian_Eff}
  \hat{H}_{\mathrm{eff}}
  &= -\widetilde{\Delta}_{c}\ha^{\dag}\ha
    +\sum_{j}(E_{j}-E_{0})\hb_{j}^{\dag}\hb_{j} \nonumber \\
  & \hspace{3em}+
    (\ha^{\dag}+\ha)
    \sum_{j}\left(\nu_{j}\hb_{j}^{\dag}\hb_{0}+h.c.\right),
\end{align}
where the effective cavity detuning is defined as $\widetilde{\Delta}_{c} = \Delta_{c}-\sgn(\Delta_{a})U_{0}N\langle\Psi^{(1)}(0)| \cos^{2}(\bk_{c}\cdot\br)
|\Psi^{(1)}(0)\rangle$, and here we set $U_{0}=g^{2}/|\Delta_{a}|$. The scattering matrix elements are given by
$\nu_{j}=\sgn(\Delta_{a})\sqrt{U_{0}}\ \theta_{j}$ with
$\theta_{j}=\langle\Psi^{(j)}(\bk_{1})|\hat{\Theta}|\Psi^{(1)}(0)\rangle$. $E_{0}$ and $E_{j}$ are eigenvalues of $\hat{\mathcal{H}}_{0}$
corresponding to $|\Psi^{(1)}(0)\rangle$ and $|\Psi^{(j)}(\bk_{1})\rangle$, respectively.

We proceed with mean-field description by taking the average values $\langle\ha\rangle=\alpha$, $\langle\hb_{0}\rangle=\psi_{0}$, and $\langle\hb_{j}\rangle=\psi_{j}$.
The atomic fields are governed by Heisenberg equations $i\partial_{t}\hb_{j}=[\hb_{j},\hat{H}_{\mathrm{eff}}]$. We seek a steady state in which
$\partial_{t}\psi_{j}=0$, gaining $\psi_{j} = (\alpha^{\ast}+\alpha)\nu_{j}\psi_{0}/(E_{0}-E_{j})$.
Together with the normalization condition $|\psi_{0}|^{2}=N-\sum_{j}|\psi_{j}|^{2}$, we retain
the ground state energy up to the forth order in amplitude of $\alpha$,
\begin{align}
  \hspace{-0.5em}
  \mathcal{E}_{\alpha}
  &\approx -\widetilde{\Delta}_{c} \alpha^{\ast}\alpha
    - NU_{0}\chi(\alpha^{\ast}+\alpha)^{2}
    + NU_{0}^{2}\chi\eta (\alpha^{\ast}+\alpha)^{4},
\end{align}
here we have defined the susceptibility $\chi = \sum_{j} |\theta_{j}|^{2}/(E_{j}-E_{0})$, and $\eta = \sum_{j} |\theta_{j}|^{2}/(E_{j}-E_{0})^{2}$.
Note that $\chi$ and $\eta$ are function of $V_{1}$, $V_{2}$, and they also depend on $\sgn(\Delta_{a})$.
Meanwhile, the Heisenberg equation of photon operator is given by $i\partial_{t}\ha=[\ha,\hat{H}_{\mathrm{eff}}]-i\kappa\ha$.
The steady condition $\partial_{t}\alpha=0$ leads to $\alpha =\sgn(\Delta_{a})\sqrt{U_{0}}\Theta/(\widetilde{\Delta}_{c}+i\kappa)$,
where an order parameter $\Theta=\int d\br \langle\hPsi^{\dag}\hat{\Theta}\hPsi\rangle$ has been introduced, so that energy can be transformed into
\begin{align}
  \label{eq:E(Theta)}
  \hspace{-0.3em}
  \mathcal{E}_{\Theta}
  &\approx -\frac{\widetilde{\Delta}_{c}U_{0}}
    {\widetilde{\Delta}_{c}^{2}+\kappa^{2}} \left(
    1 + \frac{4\widetilde{\Delta}_{c}NU_{0}\chi}
    {\widetilde{\Delta}_{c}^{2}+\kappa^{2}}
     \right) \Theta^{2} +
    \frac{16\widetilde{\Delta}_{c}^{4}NU_{0}^{4}\chi\eta}
    {(\widetilde{\Delta}_{c}^{2}+\kappa^{2})^{4}} \Theta^{4}.
\end{align}
The coefficient of the forth-order term being positive guarantees the stability of system.
We minimize the ground state energy Eq.~(\ref{eq:E(Theta)}) with respect to the order parameter $\Theta$, yielding the established superradiant phase condition
\begin{align}
  \label{eq:PhsCond}
  -\frac{4\widetilde{\Delta}_{c}NU_{0}}
  {\widetilde{\Delta}_{c}^{2}+\kappa^{2}} \chi>1.
\end{align}
Note that we only deal with $\widetilde{\Delta}_{c}<0$ here. In the normal phase, we have $\Theta=0$ as expected; while in the superradiance, system has to choose among the two solutions
\begin{align}
  \Theta
  = \pm \frac{N\eta\chi(\widetilde{\Delta}^{2}+\kappa^{2})
  \sqrt{\widetilde{\Delta}^{2}+\kappa^{2} +
  4NU_{0}\widetilde{\Delta}_{c}\chi}}
  {4\sqrt{2}(NU_{0}\widetilde{\Delta}_{c}\eta\chi)^{3/2}}.
\end{align}
In fact, the effective Hamiltonian Eq.~(\ref{eq:Hamiltonian_Eff}) possesses a $\mathbb{Z}_{2}$ symmetry, as it is invariant under simultaneous transformation $\ha \rightarrow -\ha$ and
$\hb_{j} \rightarrow -\hb_{j}$. The system spontaneously breaks this $\mathbb{Z}_{2}$ symmetry during the transition from the normal superfluid to the
superradiance~\cite{ETH17}.

\section{Results and Discussion}

The key part of phase condition Eq.~(\ref{eq:PhsCond}) is the susceptibility $\chi$ of the normal phase, which characterizes the
tendency of inducing superradiance. A larger $\chi$ indicates a greater critical magnitude of effective cavity detuning $|\widetilde{\Delta}_{c}|$.
Since we have taken the interference between two pumping lasers into account, the relations of $\chi$ to $V_{1}$ and $V_{2}$ can not be separated independently, i.e.,
$\chi(V_{1},V_{2})\neq\chi_{1}(V_{1})\chi_{2}(V_{2})$, so it is infeasible to directly apply outcomes derived from single-pump case for simplicity.

\begin{figure}[!htbp]
  \centering
  \subfloat
  {\label{fig:chi_ND} \label{fig:chi_PD}
    \includegraphics[width=0.47\textwidth]{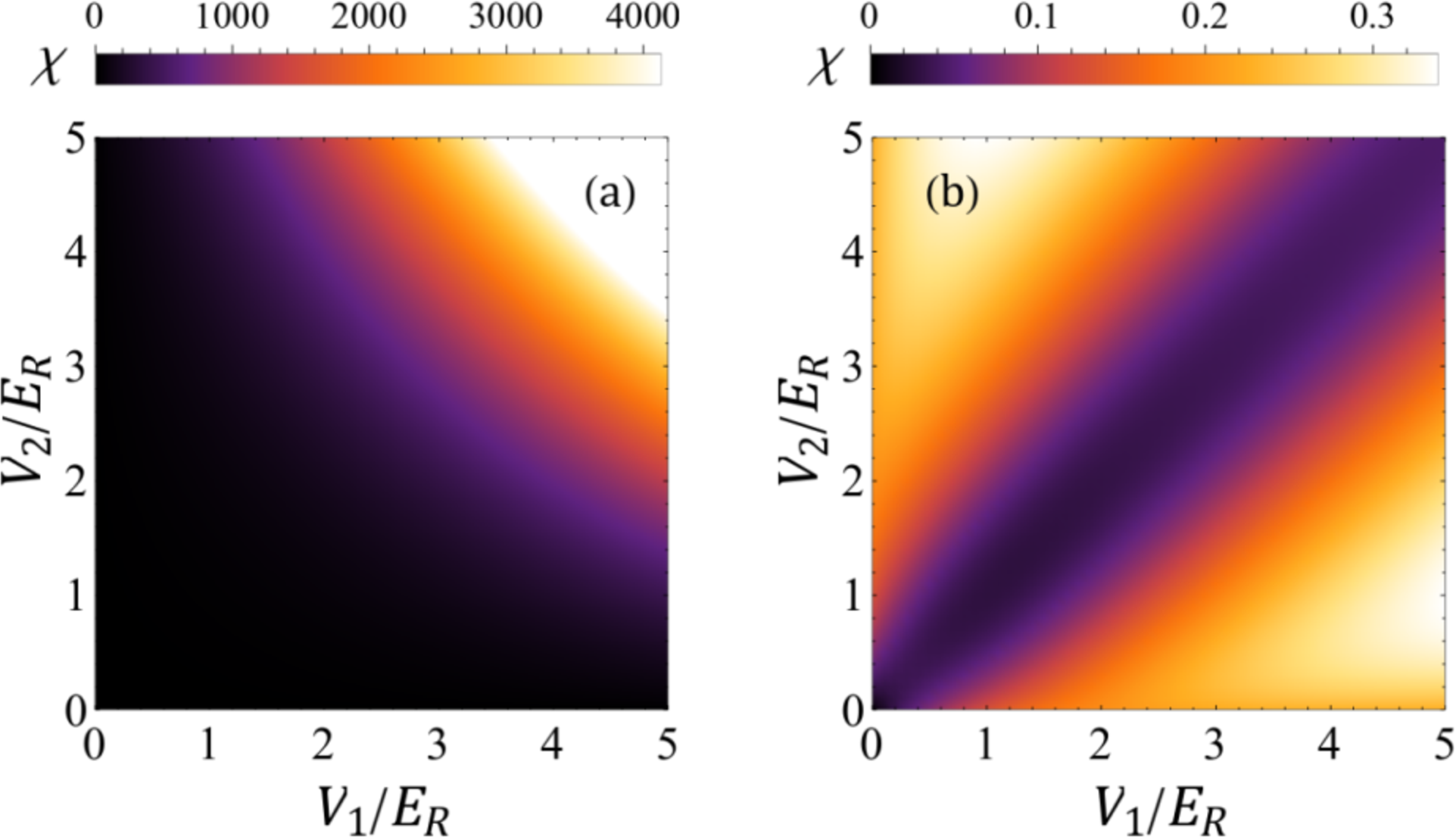}
  }
  \caption{Susceptibility $\chi$ as a function of $V_{1}$ and $V_{2}$
    for (a) $\Delta_{a}<0$ and (b) $\Delta_{a}>0$. The brightness of the region is proportional to the magnitude of $\chi$.
    A brighter region represents a larger $\chi$.}
  \label{fig:chi}
\end{figure}
\begin{figure}[!htbp]
  \centering
  \includegraphics[width=0.4\textwidth]{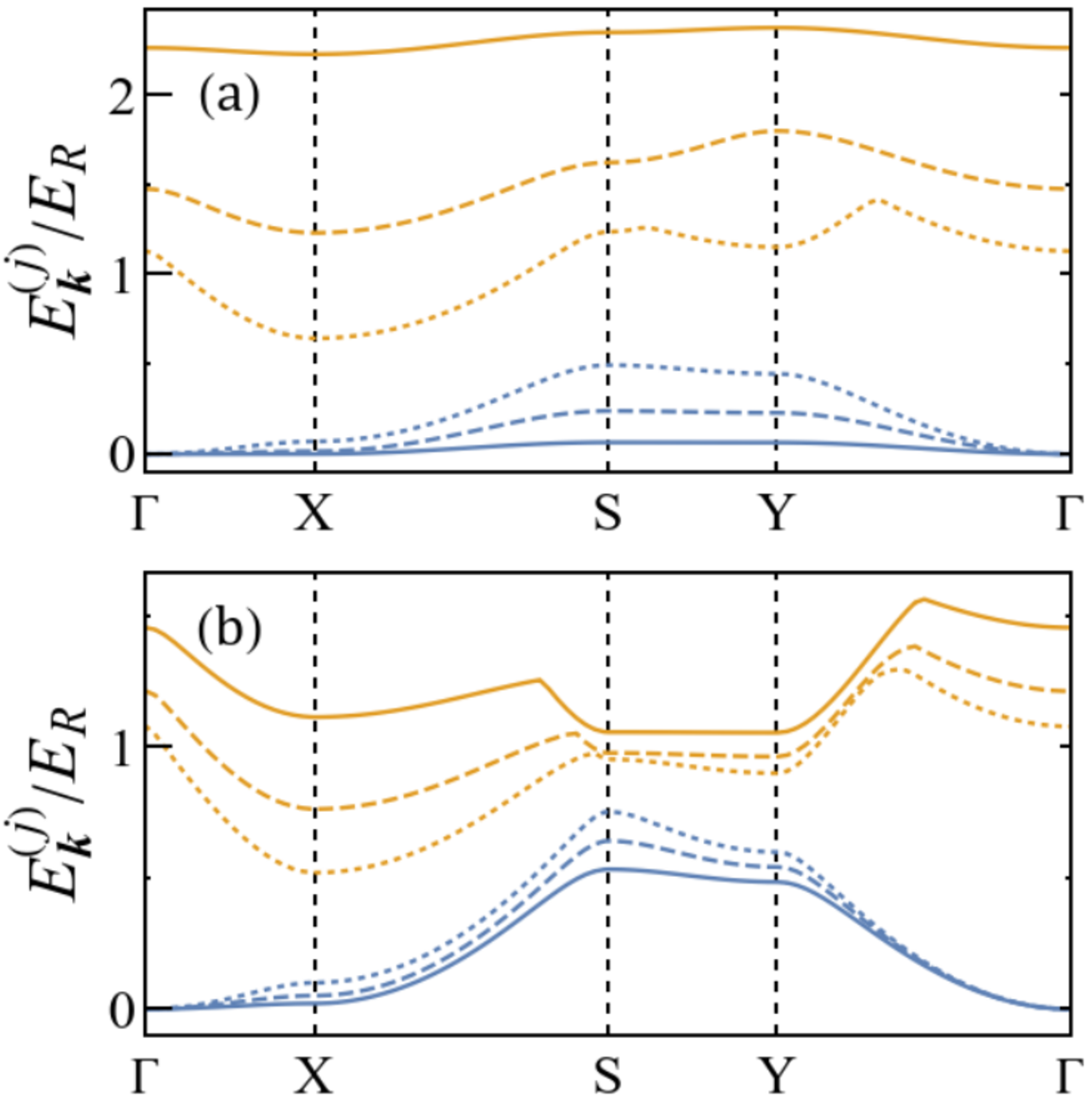}
  \caption{Lowest two bands of rectangular pump lattice for equal pump
    depths, $V_{1}=V_{2}$, with (a) $\Delta_{a}<0$ and (b)
    $\Delta_{a}>0$.
    Dotted, dashed and solid lines are numerical results of pump
    lattice depths $0.5E_{R}$, $E_{R}$ and $2E_{R}$, respectively.
    Bands are calculated along points $\Gamma$, $X$, $S$ and $Y$
    marked in
    FIG.~\ref{fig:Schematic_Diagrams}\protect\subref{fig:K_Dist}.}
  \label{fig:BandStruc}
\end{figure}

We present numerical results for $\chi$ at zero temperature in FIG.~$\ref{fig:chi}$ for different sign of atomic detuning $\Delta_{a}$.
In both cases, the susceptibility shows non-negativity $\chi \geqslant 0$. As expected, it is symmetric about $V_{1}=V_{2}$, which renders that it is equivalent to fix one of the pump depths while varying
another. However, in the case of $\Delta_{a}<0$, $\chi$ grows as $V_{1}$ increases at given $V_{2}$; while for $\Delta_{a}>0$, this monotonicity disappears as $V_{2}$ rises, and $\chi$ decreases to a local minimum towards $V_{1}=V_{2}$. Furthermore, the growth rate of $\chi$ connected with $\Delta_{a}<0$ is much larger than that with $\Delta_{a}>0$.
Within the range of $ V_{1},V_{2} \in [0,5E_{R}]$, the maximum $\chi_{\mathrm{max}}$ is over $4000$ in FIG.~$\ref{fig:chi}\subref{fig:chi_ND}$, in comparison to $\chi_{\mathrm{max}}\sim 0.34$ in
FIG.~$\ref{fig:chi}\subref{fig:chi_PD}$. This can be understood from the fact that the pump depths flatten the energy bands, as indicated in FIG.~$\ref{fig:BandStruc}$.
Here, we only plot two lowest bands, which contribute the major parts of $\chi$. Without loss of generality, pump depths are set to be equal, so that pump potentials form rectangular lattice, as shown in FIG.~$\ref{fig:Schematic_Diagrams}\subref{fig:R_Dist_NP}$. It is clear that increasing pump depths has a much more powerful effect on flattening energy bands and widening band gaps with
negativity than positivity of $\Delta_{a}$. Especially the first band width $E_{1}-E_{0}$, which determines the denominator of the leading term in $\chi$, becomes dominant at higher pump depths.

With susceptibility $\chi$ in hand, we are now in a position to construct phase diagrams. The pump lattice depths $V_{1}$, $V_{2}$ can be tuned by varying the pump laser power, while cavity detuning
$\Delta_{c}$ can be controlled via cavity frequency $\omega_{c}$. For experimental consideration, we shall construct phase diagrams in terms of experimentally tunable parameters $V_{1}$, $V_{2}$ and $\Delta_{c}$.

\subsection{Phase Diagrams Spanned by $V_{1}$ and $\Delta_{c}$}

Solving phase boundary condition of Eq.~(\ref{eq:PhsCond}) with respect to $\Delta_{c}$, yields the critical cavity detuning
\begin{align}
  \frac{\Delta_{\pm}}{NU_{0}}
  =\sgn(\Delta_{a})\theta_{c} -2\chi
  \pm \sqrt{4\chi^{2}- \left(\frac{\kappa}{NU_{0}}\right)^{2}},
  \label{eq:Delta_cr}
\end{align}
where we additionally define scattering matrix element $\theta_{c} = \langle\Psi^{(1)}(0)| \cos^{2}(\bk_{c}\cdot\br)|\Psi^{(1)}(0)\rangle$ for convenience. It turns out that $\theta_{c}$ can not be neglected when $\Delta_{a}>0$, as $\theta_{c}$ is comparable with $\chi$.

\begin{figure}[!htbp]
  \centering
  \subfloat
  {\label{fig:PhsDgm_Delta_V1_sV2_ND} \label{fig:PhsDgm_Delta_V1_sV2_PD}
    \includegraphics[width=0.47\textwidth]{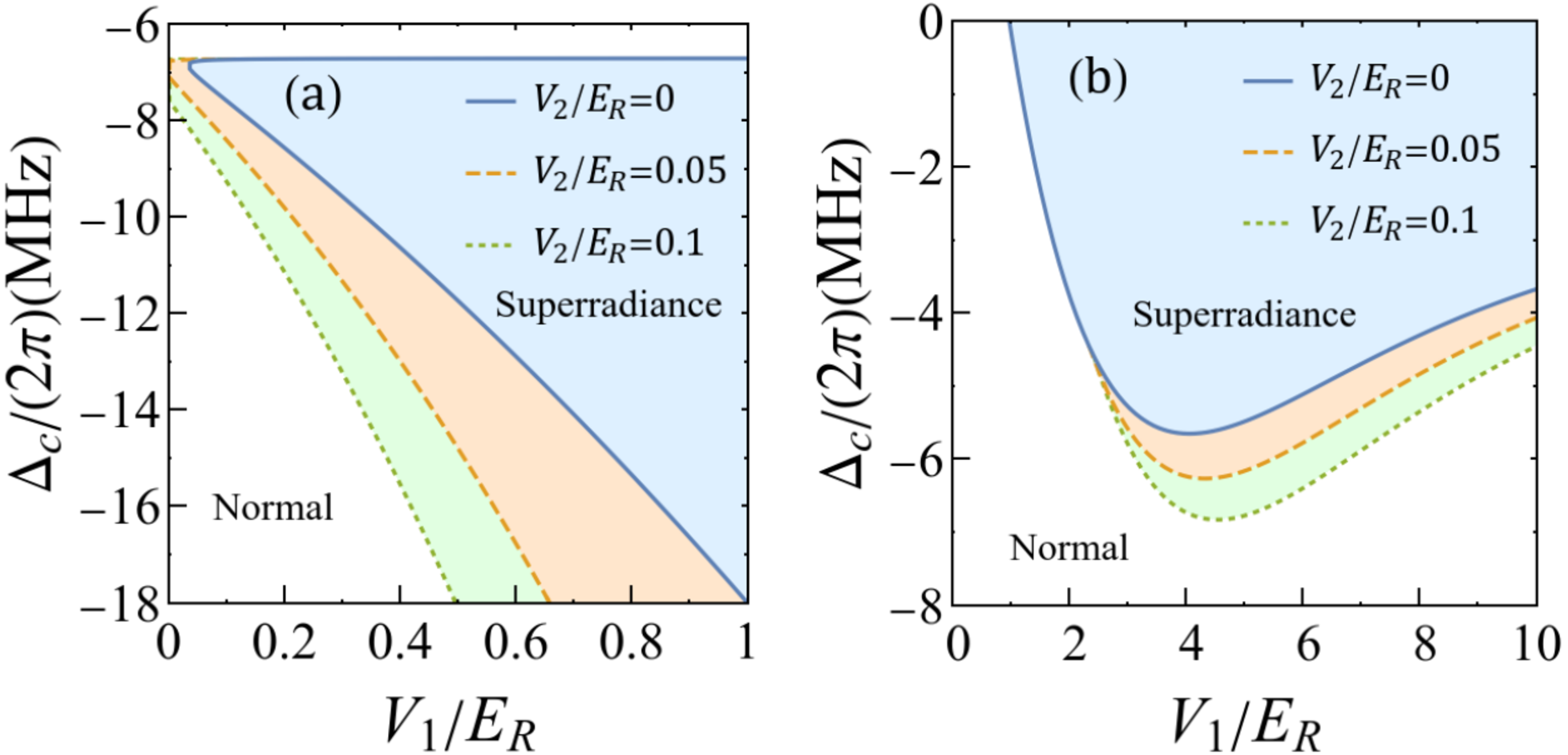}
  }
  \caption{Phase diagrams spanned by one of pump lattice depths
    $V_{1}$ and cavity detuning $\Delta_{c}$ for (a) $\Delta_{a}<0$
    and (b) $\Delta_{a}>0$.
    Another pump depth is fixed at small value, $V_{2}/E_{R}<1$.
    Experimental settings in Ref.~\cite{DON19} are
    used, where atoms number $N=2.7 \times 10^{5}$,
    decay rate $\kappa = 2\pi \times 147\ \mathrm{kHz}$,
    atom-cavity coupling $g = 2\pi \times 1.95\ \mathrm{MHz}$ and
    magnitude of atomic detuning $|\Delta_{a}| = 2\pi \times 76.6\
    \mathrm{GHz}$.}
  \label{fig:PhsDgm_Delta_V1_sV2}
\end{figure}

Here, we only consider the situation that $\Delta_{c}<0$. The reason is that: for $\Delta_{c}>0$, the term $-\Delta_{c}\ha^{\dag}\ha$ in Eq.~(\ref{eq:Hamiltonian_Origin}) indicates that the system can lower its energy in the rotating frame with more photons, and then gives rise to a transition from metastable regime to unstable regime~\cite{DON19}. In this situation, we have to care about large $\alpha$, and therefore, perturbation method breaks down. Instead, within $\Delta_{c}<0$, we only need to deal with small $\alpha$, and the above results always hold.
Hence, only negative solutions of Eq.~(\ref{eq:Delta_cr}) have been taken into account. We map out phase diagrams on $V_{1}$ and $\Delta_{c}$ at small $V_{2}$ in FIG.~$\ref{fig:PhsDgm_Delta_V1_sV2}$.
Here, we choose the parameters from Ref.~\cite{DON19} for verification, where atoms number $N=2.7 \times 10^{5}$, decay rate $\kappa = 2\pi \times 147\ \mathrm{kHz}$, atom-cavity coupling
$g = 2\pi \times 1.95\ \mathrm{MHz}$ and magnitude of atomic detuning $|\Delta_{a}| = 2\pi \times 76.6\ \mathrm{GHz}$.

For $\Delta_{a}<0$, $\Delta_{\pm}$ are both negative due to $\chi>0$ and $\theta_{c}>0$. At the limit $V_{2} \rightarrow 0$, phase boundary in
FIG.~$\ref{fig:PhsDgm_Delta_V1_sV2}\subref{fig:PhsDgm_Delta_V1_sV2_ND}$ captures the characteristics of the well-known superradiant transition
in single-pump system: the cavity photons emerge and then vanish as the cavity detuning increasing from below~\cite{ESS10}.
And there is a threhold of $V_{1}$, here $V_{1} \sim 0.03E_{R}$, below which the superradiance will not happen. The threhold can be obtained
by requiring $\chi=\kappa/(2NU_{0})$. However, this threshold will no longer exist while another pump depth $V_{2}$ rises, as exhibited by the boundaries of $0.05E_{R}$ and
$0.1E_{R}$ in FIG.~$\ref{fig:PhsDgm_Delta_V1_sV2}\subref{fig:PhsDgm_Delta_V1_sV2_ND}$. In other words, if a system is in the superradiant phase with one pump laser, adding another pump will not drive it out. And increasing $V_{2}$ also makes the lower bounds of phase boundaries steeper and steeper, and indicates the system is easier to enter superradiant phase.

However, situations are different in the case of $\Delta_{a}>0$, where $\theta_{c}\sim 0.5$ is comparable with $\chi$. And that leaves only $\Delta_{-}$ being partially negative as a
function of $V_{1}$ at current parameter settings. At the limit $V_{2} \rightarrow 0$, phase boundary in FIG.~$\ref{fig:PhsDgm_Delta_V1_sV2}\subref{fig:PhsDgm_Delta_V1_sV2_PD}$
is in agreement with the numerical mean-field results in Ref.~\cite{DON19}. For a fixed cavity detuning $\Delta_{c}$, such as $-2\pi \times 4\ \mathrm{MHz}$,
the system starting in the normal phase undergoes a transition to the superradiant phase and comes back to the normal phase as pump lattice is ramped up.
This uniqueness of the phase boundary is a result of the competition between band gap and mode softening. Meanwhile, there is a local minimum of critical $\Delta_{c}$ for
different $V_{2}$, and it gets lower as $V_{2}$ increases.

\begin{figure}[!htbp]
  \centering
  \subfloat
  {\label{fig:PhsDgm_Delta_V1_lV2_ND} \label{fig:PhsDgm_Delta_V1_lV2_PD}
    \includegraphics[width=0.47\textwidth]{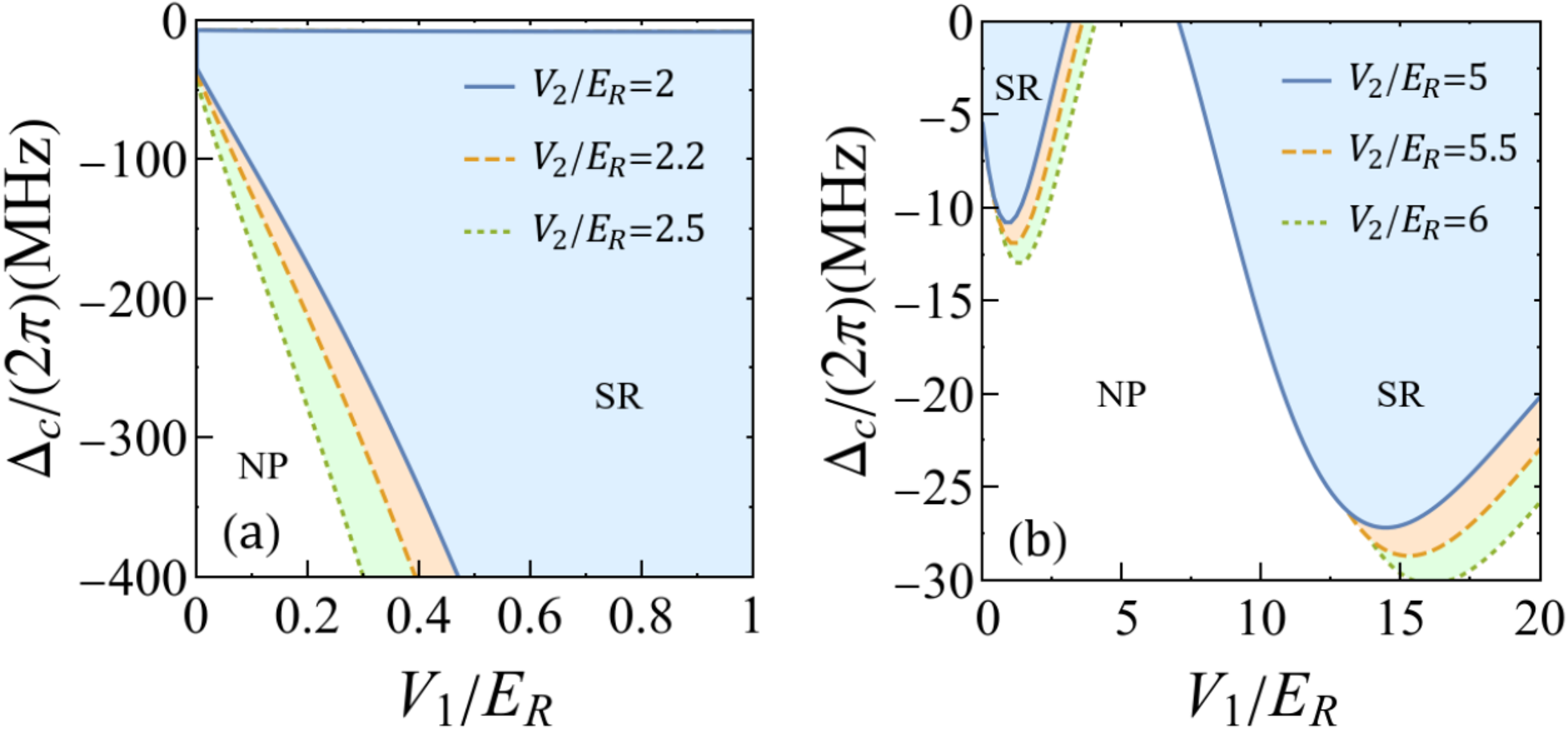}
  }
  \caption{Phase diagrams spanned by one of pump lattice depths
    $V_{1}$ and cavity detuning $\Delta_{c}$ for (a) $\Delta_{a}<0$
    and (b) $\Delta_{a}>0$.
    Another pump depth is fixed at comparable value, $V_{2}/E_{R}>1$.
    Other parameters are set to be the same as
    FIG~$\ref{fig:PhsDgm_Delta_V1_sV2}$.
    The acronym NP stands for the Normal Phase, and SR for
    Superradiance.}
  \label{fig:PhsDgm_Delta_V1_lV2}
\end{figure}

In the regime of a deeper pump lattice, $V_{2}/E_{R}>1$, the slope of critical cavity detuning $\Delta_{-}$, as a function of $V_{1}$, gets even more steeper than in $V_{2}/E_{R}<1$ for
negative $\Delta_{a}$, as shown in FIG.~$\ref{fig:PhsDgm_Delta_V1_lV2}\subref{fig:PhsDgm_Delta_V1_lV2_ND}$. It indicates the increasing tendency of transition towards
superradiance as the pump depth goes deeper, provided that the pump lattice potentials are both attractive.

While for positive $\Delta_{a}$, another superradiant region shows up at higher pump depths, $V_{2}/E_{R}>1$. In this regime, the system enters the superradiant phase, comes back
to the normal phase, and then reenters into the superradiance as pump depth $V_{1}$ is ramped up, as illustrated in FIG.~$\ref{fig:PhsDgm_Delta_V1_lV2}\subref{fig:PhsDgm_Delta_V1_lV2_PD}$.
In contrast to the case in $\Delta_{a}<0$, if a system is in the superradiant phase with single pump laser, adding another pump will suppress superradiance within a certain range.

\subsection{Phase Diagrams Spanned by $V_{1}$ and $V_{2}$}
\begin{figure}[!htbp]
  \centering
  \subfloat
  {\label{fig:PhsDgm_V1_V2_ND} \label{fig:PhsDgm_V1_V2_PD}
    \includegraphics[width=0.47\textwidth]{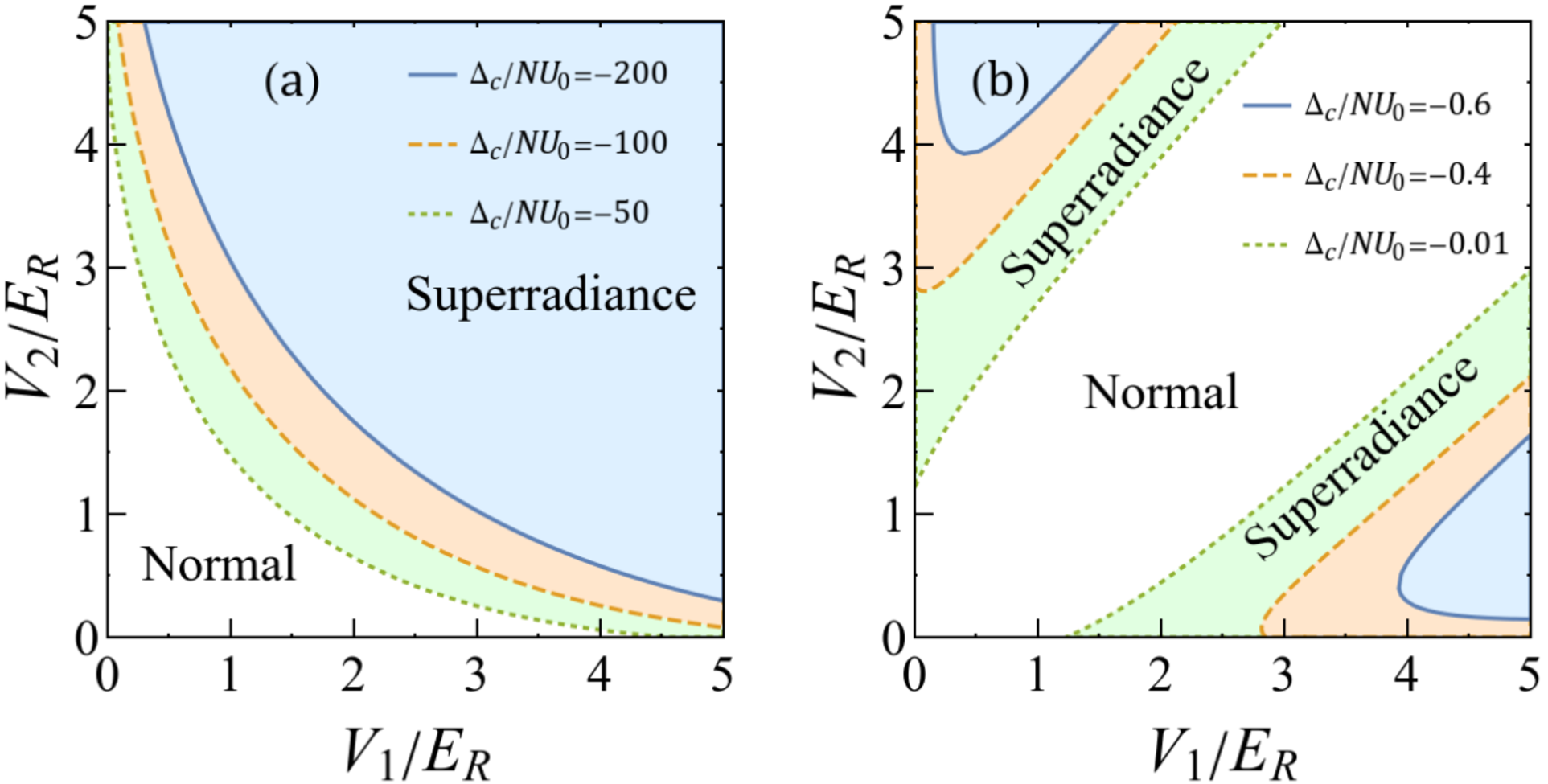}
  }
  \caption{Phase diagrams in terms of two pump lattice depths $V_{1}$
    and $V_{2}$ with cavity detuning $\Delta_{c}$ fixed for (a)
    $\Delta_{a}<0$ and (b) $\Delta_{a}>0$.
    Here we set $\kappa/(NU_{0}) = 0.1$.}
  \label{fig:PhsDgm_V1_V2}
\end{figure}

In phase boundary condition of Eq.~(\ref{eq:PhsCond}), not only $\chi$, but also $\theta_{c}$ in $\widetilde{\Delta}_{c}$ has the dependence on $V_{1}$ and $V_{2}$.
It is useful to introduce functions $f_{\pm}$ for $\sgn(\Delta_{a}) = \pm 1$, where
\begin{align}
  f_{\pm}
  =
  \chi
  + \frac{[\Delta_{c}/(NU_{0})\mp\theta_{c}]^{2}
  + [\kappa/(NU_{0})]^{2}}
  {4[\Delta_{c}/(NU_{0})\mp\theta_{c}]}.
\end{align}
Note that $f_{\pm}$ are dimensionless functions of $V_{1}/E_{R}$ and $V_{2}/E_{R}$ if $\Delta_{c}/(NU_{0})$ and $\kappa/(NU_{0})$ are given.
$\widetilde{\Delta}_{c}<0$ and $\Delta_{c}<0$ restrict $\Delta_{c}/(NU_{0})<0$ for $\Delta_{a}>0$ and $\Delta_{c}/(NU_{0})<-\theta_{c}$ for $\Delta_{a}<0$.
Here, $f_{\pm}>0$ are the criteria for superradiance, and $f_{\pm}<0$ for the normal phase. In FIG.~$\ref{fig:PhsDgm_V1_V2}$, we plot the phase diagrams for
different cavity detunings $\Delta_{c}/(NU_{0})$ at decay rate $\kappa/(NU_{0}) = 0.1$.

Within $V_{1},V_{2} \leqslant 5E_{R}$, the area of the normal phase shrinks as $\Delta_{c}$ approaches $0$ for both positive and negative $\Delta_{a}$, in accordance with the above results.
Now, if we keep two pump depths equal, $V_{1}=V_{2}$, and increase them simultaneously, the system undergoes a transition from the normal phase to superradiance in the case of positive
$\Delta_{a}$; while for negative $\Delta_{a}$, the system remains in the normal phase. Notice that there exists a diagonal region around $V_{1}=V_{2}$ for the normal phase when $\Delta_{a}>0$, even if $\Delta_{c}$ approaches zero, as shown in FIG.~$\ref{fig:PhsDgm_V1_V2}\subref{fig:PhsDgm_V1_V2_PD}$. Actually, one can derive that $f_{+}<\chi-\theta_{c}/4$ is true for
arbitrary $V_{1},V_{2}>0$ in this case. Numerical results show that $\chi-\theta_{c}/4 \sim -0.1$, thereby $f_{+}<0$, as two equal pump depths $V_{1}=V_{2}=V$ change within $10E_{R}$.
In other words, if two pump depths are equal, the system will remain in the normal phase as pump depths increases within a wide range (at least from $0$ to $10E_{R}$), no matter how we choose $\kappa$ and $\Delta_{c}$. The same arguments apply to the neighborhoods of $V_{1}=V_{2}$. From another point of view, if a system is in the superradiant phase with single pump of depth $V_{1}$, adding another pump of $V_{2}$ will suppress superradiance for $V_{2} \leqslant V_{1}$.

\section{Conclusion}

We compare susceptibilities of a BEC inside two crossed pump fields for the red atomic detuning ($\Delta_{a}<0$) and the blue atomic detuning ($\Delta_{a}>0$).
We map out phase boundaries separating the normal superfluid and the superradiance for the red cavity detuning ($\Delta_{c}<0$). Phase diagrams are examined at the single pump limit, and they are
also compared in the double pump regime. We find that if a system is in the superradiant phase with one pump laser, adding another pump will remain in the superradiance
when the atomic detuning is red ($\Delta_{a}<0$). While for the blue detuning ($\Delta_{a}>0$), increasing another pump potential have a suppressive effect on the superradiance.
Finally, we show that equally increasing two pump potentials can also induce a transition from the normal phase to superradiance for red atomic detuning ($\Delta_{a}<0$).
While in the case of blue detuning ($\Delta_{a}>0$), the system will remain in the normal phase as pump depths increasing within a wide range, and it is independent of cavity detuning and decay rate.
Experimental verifications of our predictions are believed to contribute to a better understanding of superradiance transitions with utracold gases in optical cavities.

\section*{acknowledgments}
This work is supported by NSFC under Grants No.12174055 and No.11674058, and by the Natural Science
Foundation of Fujian under Grant No. 2020J01195.
%

\end{document}